# Unsupervised Tissue Segmentation via Deep Constrained Gaussian Network

Yang Nan, Peng Tang, Guyue Zhang, Caihong Zeng, Zhihong Liu, Zhifan Gao, Heye Zhang, Guang Yang, *Senior Member, IEEE*

***Abstract*—Tissue segmentation is the mainstay of pathological examination, whereas the manual delineation is unduly burdensome. To assist this time-consuming and subjective manual step, researchers have devised methods to automatically segment structures in pathological images. Recently, automated machine and deep learning based methods dominate tissue segmentation research studies. However, most machine and deep learning based approaches are supervised and developed using a large number of training samples, in which the pixel-wise annotations are expensive and sometimes can be impossible to obtain. This paper introduces a novel unsupervised learning paradigm by integrating an end-to-end deep mixture model with a constrained indicator to acquire accurate semantic tissue segmentation. This constraint aims to centralise the components of deep mixture models during the calculation of the optimisation function. In so doing, the redundant or empty class issues, which are common in current unsupervised learning methods, can be greatly reduced. By validation on both public and in-house datasets, the proposed deep constrained Gaussian network achieves significantly (Wilcoxon signed-rank test) better performance (with the average Dice scores of 0.737 and 0.735, respectively) on tissue segmentation with improved stability and robustness, compared to other existing unsupervised segmentation approaches. Furthermore, the proposed method presents a similar performance (p-value > 0.05) compared to the fully supervised U-Net.**

## I. Introduction

Given an image, a segmentation algorithm aims to assign labels for pixels based on their feature representations. Tissue segmentation is essential for automated pathological examination, diagnosis and prognosis; however, manual delineation is time-consuming, onerous and unreproducible. To alleviate the burden of this manual procedure, researchers have explored conventional approaches to automatically segment organs or structures, including watershed [1], contour detection [2], clustering [3, 4], and random field [5], etc. However, these methods are unreliable and heavily rely on thresholds or preset parameters. Recently, machine and deep learning based methods have garnered great success in computational pathology [6-9]. For example, Mahbod et al.

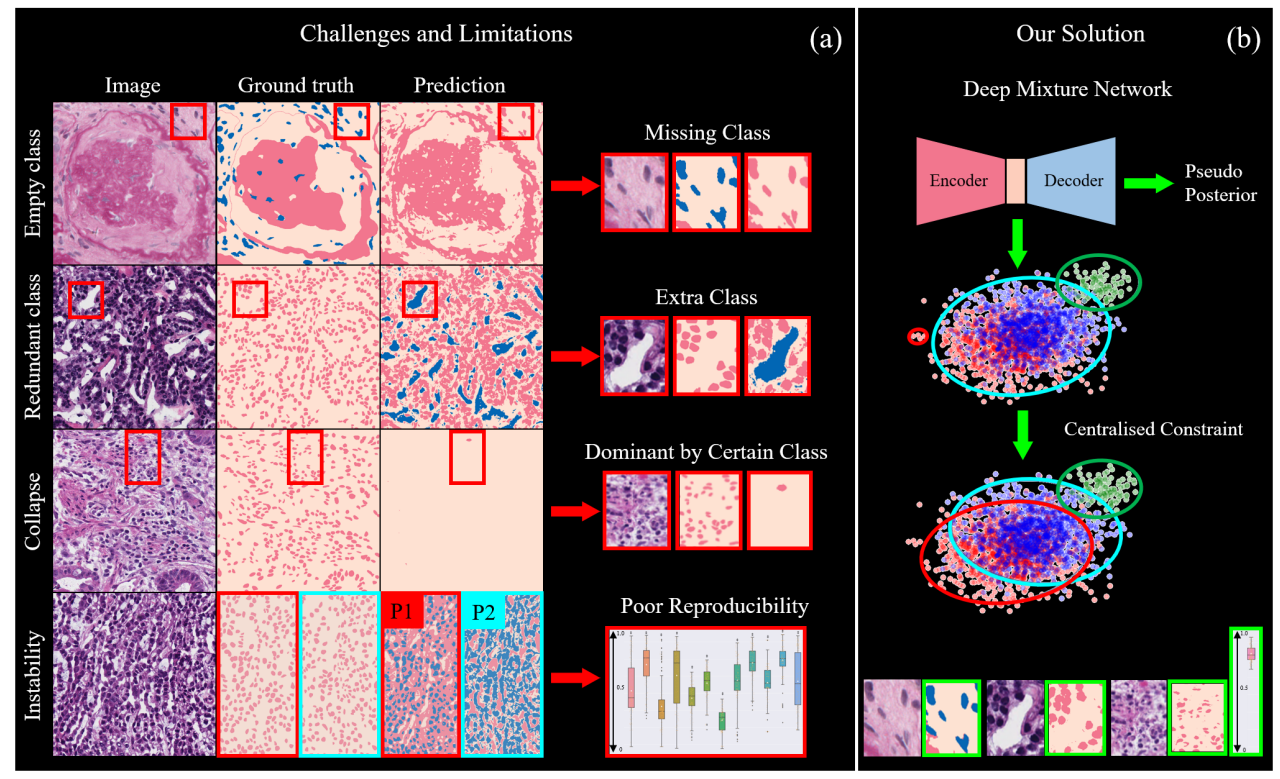

Fig. 1. Current challenges and limitations of unsupervised segmentation for tissue segmentation (a) and our solutions (b). (a) examples of empty class (first row), redundant class (second row), collapse (third row), and instability (fourth row) issues. The red boxes highlight three subregions of the raw image, ground truth and prediction (from left to right) using existing unsupervised segmentation methods. P1 and P2 represent the first and second predictions obtained from repeated experimental studies (last row); (b) our proposed unsupervised segmentation based on a centralised constraint deep mixture network. The representative results of our proposed model are highlighted in green boxes (last row), and from left to right, these show clearly that our unsupervised segmentation can tackle empty class, redundant class, collapse, and instability issues. All box plot scales range from [0, 1] for the Dice scores.

[9] proposed a progressive sequential causal GAN to synthesize the late gadolinium enhancement imaging for better segmentation of diagnosis-related structures. Liu et al. [10] incorporated CycleGAN with an adaptive Mask RCNN for unsupervised nuclei segmentation in histopathology images, by learning knowledge from fluorescence microscopy images. However, most learning-based methods are fully supervised which require manual labelling, or unsupervised that demand complex training procedures. In particular, complex pathological structures dramatically increase the difficulty of pixel-level annotation, resulting in an urgent need for developing segmentation methods with limited or no manual annotation.

One way to overcome this hurdle is known as (deep) semi-supervised learning, which builds the model with limited

Yang Nan and Guang Yang are with National Heart and Lung Institute, Imperial College London, London, UK (Send correspondence to: {y.nan20,g.yang@imperial.ac.uk}, zhangheye@mail.sysu.edu.cn).
Peng Tang is with Department of Informatics, Technical University of Munich.
Guyue Zhang is with Zhejiang Institute of Standardization, Hangzhou, China.
Caihong Zeng and Zhihong Liu are with National Clinical Research Centre of Kidney Diseases, Nanjing University, Nanjing, China.
Zhifan Gao and Heye Zhang are with School of Biomedical Engineering Sun Yat-sen University, China.

annotations or prior knowledge of the targets. Self-training is a commonly used method that trains the model with limited annotated labels and fine-tunes it via pseudo labels generated by itself. For instance, Liang et al. [11] proposed an iterative learning scheme to segment gastric tumours based on a partially labelled dataset. In addition to self-training, one can use the prior knowledge given by conventional methods or empirical constraints such as target labels to train a network. This includes the utilization of coarse masks given by image processing algorithms, pre-trained weights from correlated datasets, or image-level annotations provided by domain experts. Hu et al. [12] applied activation maps to detect COVID-19 infections without pixel-level annotation. Atlason et al. [13] took coarse masks from an automated labelling system as attention maps to force the network to concentrate on the constrained region.

Another solution is (deep) unsupervised learning, which produces general semantic predictions such as ‘background’ and ‘foreground’ without using any manual annotations. For instance, Kanezaki et al. [14] employed Simple Linear Iterative Clustering [15] to obtain super-pixel level segmentation results, combining with convolutional neural networks to segment natural images. Shen et al. [16] introduced a coupled “deep-image-prior” module to segment background and foreground regions. However, most of these studies focused on natural images, whose effectiveness for pathological images remains unclear. Moreover, image quality variations (e.g., different brightness, contrast, noise, and shade levels in pathological images) may lead to poor generalisability for models originally developed for natural images. The randomized initializations of some unsupervised learning methods may further result in unreliable performance and weak reproducibility. In particular, there are several degenerative issues (Fig. 1) for unsupervised segmentation, including (1) empty class (2) redundant class (3) collapse, and (4) instability issues. The empty class problem indicates that the model confounds a certain class with another one, e.g., the prediction only has two classes even if the pre-defined number of classes is three (Fig. 1 (a) first row). The redundant class indicates the demand for an additional class to achieve better performance during unsupervised segmentation. This redundant class is used to represent the hard samples, which are defined as pixels whose intensities are diffusely/narrowly vary from the average intensity of their true/false class. For example, the white regions in the second row of Fig. 1 (a) are considered a unique class, since the model cannot treat them as the same class (background) as stroma. Collapse issue refers to the phenomenon when a certain class dominates the major predictions of an image while other classes only appear sporadically (as shown in Fig. 1(a), the third row). The instability means the fluctuant performance when conducting repeated training (Fig. 1 (a) fourth row).

To address these limitations, our study proposes a novel unsupervised approach that integrates a deep neural network with log-likelihood maximisation and centralised constraint (Fig. 1 (b)), namely Deep Constrained Gaussian Network (dubbed DCGN). Unlike previous methods that utilise prior knowledge, the proposed DCGN takes raw images as inputs and produces pixel-wise predictions for tissue structures. Besides, a centralised constraint, which can greatly enhance the model’s robustness and performance, is devised, aiming to shrink the estimated mean value of the components closer to the real data centroids. Comprehensive experimental studies were conducted on a multicentre open access dataset (i.e., MoNuSeg, acquired from the TCGA archive) and our in-house dataset. In addition, repeated experiments are performed to evaluate the stability of different approaches. The proposed method achieves a new state-of-the-art performance in unsupervised segmentation in pathological images, with Dice scores of 0.743 and 0.737 on MoNuSeg and our in-house dataset, respectively, outperforming all comparison models significantly (Wilcoxon signed-rank test p-value<0.001). The main contributions of this paper are:

1) Major challenges and limitations of current unsupervised tissue segmentation approaches in the pathological image domain have been investigated comprehensively and summarised concisely. These include the missing class problem, the redundant class problem, collapse, and the instability issues. We observed that these degenerative issues are caused by large intra-class variations or small inter-class variations.

2) A DCGN with a centralised constraint is proposed to address all the degenerative problems. This centralised constraint forces the estimated mean to approximate the observed mean value by considering the heterogeneity of the training data to solve a) the missing class or collapse issue when previous unsupervised methods may consider outliers as a single class, b) the instability issue when previous unsupervised methods may be trapped at the local optimum, and c) the redundant class issue when the existing unsupervised methods could encounter small inter-class variations and result in weak predictions. The proposed centralised constraint is a succinct yet effective module that can be easily adapted to other unsupervised approaches for tissue segmentation.

3) Comprehensive experimental studies have been conducted to demonstrate the significantly improved performance of our proposed DCGN with greatly enhanced reproducibility. Our study also suggests that the assessment of future unsupervised tissue segmentation methods must consider degenerative problems and repeated experiments should be carried out to prove stability and robustness.

The rest of this paper is organised as follows. The related studies on unsupervised segmentation are summarised in Section II. Details of the proposed method are illustrated in Section III. The experimental settings, including dataset details and training parameters, are described in Section IV. Sections V and VI present the discussion and conclusion of this study.

## II. Related works

This section describes the most related previously published studies, including both conventional and deep learning-based unsupervised segmentation approaches.

### A. Conventional Unsupervised Segmentation

In general, unsupervised segmentation can be treated as a clustering task. Given a three-channel RGB image, the clustering algorithm first flattens the 3D array to a 2D vector, then each pixel group (pixels along with R, G, and B channels) is considered as a multidimensional sample for clustering. These methods include graph/normalised cuts [17, 18], Markov random field [18], minibatch K-means [19], Gaussian mixture model (GMM) [20], mean shift [21], and have been widely used in medical image analysis tasks, such as registration [22], lesion detection [23] and segmentation [20]. In addition to clustering, learning and distinguishing different feature representations can also segment regions of interest from images. For instance, Fan et al. [24] applied hierarchical image matting to segment vessels from fundus images. Tosun et al. [25] proposed an object-oriented method with a homogeneity measurement to segment biopsy images.

### *B. Deep Clustering and Mutual Information*

Recent studies of unsupervised learning aim to combine conventional clustering methods with deep neural networks [26-28]. Specifically, these methods use clustering-based objective functions to train a neural network. For instance, DeepCluster [26] jointly updated parameters of the neural networks and clustering during the training, and used pseudo labels to calculate objective functions. Kim et al. [29] proposed a spatial constraint to the softmax cross-entropy loss (given by pseudo labels and predictions) to keep the spatial continuity of semantic predictions. Wellmann et al. [28] integrated domain knowledge as probabilistic relations and proposed a deep conditional GMM. However, using pseudo labels for training is prone to weak solutions, such as empty clusters, and trivial parametrisation [26].

Maximizing the mutual information of paired predictions is effective [30]. To further alleviate degenerative issues, Invariant Information Clustering (IIC) [31] modified co-clustering approaches and proposed mutual information based objective functions between paired samples to train a segmentation model. Given a pair of variables $X$, $Y$ and their marginal distribution $p(x)$ and $p(y)$, the mutual information between $X$ and $Y$, jointly distributed according to $p(x, y)$, is defined
as

$$I(X;Y) = \sum_{x,y} p(x,y) \log \frac{p(x,y)}{p(x)p(y)}. \quad (1)$$

IIC generated paired images by randomised rotation to assist the network to learn the invariant information and textual representations. More generally, IIC aimed to find common parts of paired samples while ignoring the redundant ones. However, it still suffers from degenerative issues and unstable performance (as shown in Section IV).

### *C. Deep Generative Models and Log Likelihood*

Deep generative models aim to learn image representations by reconstructing the input images through generative models, such as generative adversarial networks (GAN), variational auto-encoder (VAE), and encoder-decoders. These representations can then be used to produce semantic predictions or calculate objective functions [32]. For instance, Chen et al. [33] employed redrawing ideas to segment foreground and background samples. Gandelsman et al. [34] proposed double Deep Image Prior (DIP) to composite images as background and foreground samples. However, these methods can only segment limited classes, which would be computationally redundant when producing multi-class predictions.

Another attempt is to combine deep neural networks with the GMM. Zong et al. [35] proposed a deep auto-encoder Gaussian mixture model (DAGMM), adding GMM to the low-dimensional feature representations within an auto-encoder for unsupervised anomaly detection. Oord et al. [36] incorporated GMM on the top layers in hierarchical structures for unsupervised classification. Based on these studies, Zanjani et al. [37] extended DGMM for segmentation via classifying each pixel for stain normalisation. They proposed three novel schemes, including GAN-based, VAE-based, and deep convolutional Gaussian mixture model (DCGMM) based approaches. Among these attempts, the VAE-based approach and DCGMM can be well transferred to segmentation. The VAE-based method performed log-likelihood loss and Kullback-Leibler (KL) divergence loss to assess the reconstruction performance of raw data and the correlation between latent variables and prior distribution, respectively. The DCGMM trained the network by maximising the log-likelihood objective function. However, most of these methods only simply combine expectation maximisation with deep neural networks, without addressing the common issues in unsupervised tissue segmentation.

## III. Methodology

### *A. Overview*

To address the limitations of existing unsupervised segmentation approaches, we summarise the properties that a well-performed model should possess:

1. The model should have strong reproducibility during the training and validation stages.
2. The model should be as light as possible and does not require complex pre-processing or post-processing steps.
3. The model should have the ability to alleviate degenerative issues (e.g., the empty clusters problem).

By considering the above properties, DCGN is proposed to segment pathological tissue images.

### *B. Deep Constrained Gaussian Network*

In biomedical image segmentation, especially in pathological images, the semantic labels are more related to colour representations compared to natural images. This suggests that a mixture model can be well integrated with a deep neural network for unsupervised segmentation.

Let $\omega$ denote learnable parameters of a deep neural network and $\mathcal{J}$ refer to the objective function. In fully supervised learning, $\omega$ is updated by minimizing the objective function $\mathcal{J}$, which is commonly defined by calculating the errors between ground truth labels and predictions. Therefore, the key to unsupervised segmentation can be treated as finding the best objective function for training deep neural networks without annotation (ground truth label). In addition to maximizing the mutual information between paired samples in Eq. (1), maximizing the log-likelihood can also be integrated into the gradient descent training framework, by minimizing the negative log-likelihood.

The proposed DCGN includes a feature extractor, a decoder, and a log-likelihood estimation module. Different from the accurate objective functions that calculate the error between the ground truths and predictions in supervised learning, log-likelihood maximization is a biased estimation that only produces a rough 'direction' to the global optimum [38, 39]. Therefore, we believe that complex and deep network structures are more likely to be over-fitted and trapped at local optima when there is no strong supervised optimisation function. In order to formulate a light architecture, MobileNet-V2 [40] is employed as the feature extractor, followed by a decoder that is comprised of Upsampling layers, Convolution layers, Batch normalisation layers, and ReLU activations. To adapt the prediction of the network to the pseudo posterior of the latent variable $Z$ in the mixture model, a differentiable softmax layer is applied to the output, forming a $[W, H, K]$ shaped prediction ($W$, $H$ and $K$ are the width, height, and the number of classes, respectively). Given input images $I$ with $K$ classes, the network $\emptyset$ aims to produce semantic probability maps $\varphi$, which are considered as the pseudo posterior $\gamma$ in the conventional GMM, that is

$$\gamma \approx \varphi = \emptyset(I, \omega) \in \mathbb{R}^{W \times H \times K}. \quad (2)$$

Based on the above assumption, the E-step can be conducted by forward propagation through a neural network, while M-step is applied by optimising the likelihood function via gradient descent.

Given the pseudo posterior $\gamma_{ik}$, the log-likelihood $\mathcal{L}(\Theta|\Theta^{(t)})$ of the multivariate GMM can be estimated using

$$\mathcal{L}(\Theta|\Theta^{(t)}) = \begin{matrix} \sum_{k=1}^{K}\sum_{i=1}^{N} \gamma_{ik} \, [\log\alpha_k - \frac{D}{2}\log(2\pi) - \log\gamma_{ik} - \\ \frac{1}{2}\log|\Sigma_k| - \frac{1}{2}(X_i - \mu_k)^T {\Sigma_k}^{-1}(X_i - \mu_k)]] \end{matrix}, \quad (3)$$

where $\frac{D}{2}\log(2\pi)$ is a constant that can be ignored, D is the dimension of each sample (D=3 for a flattened RGB image array), N is the number of samples (pixel groups) of the image, $\alpha_k$ is the weight of the $k$-th Gaussian mixture model that $\sum_k^K \alpha_k = 1$. Therefore, by integrating Eqs. (2) and (3), the network $\emptyset$ can be trained by minimising the log-likelihood $\mathcal{L}$

$$\omega = \arg\min_{\omega}[-\mathcal{L}(\omega)]. \quad (4)$$

It is of note that one major concern for existing deep Gaussian models is the redundant class issue, which is mainly caused by small inter-class and large intra-class variations. It makes the model assign the same (different) label(s) to samples of different (same) classes. The hard samples (outliers) may also lead to an incorrect estimate of the optimisation function, resulting in local optima trapping or an unstable training process. Another problem is the instability issue, which is a common drawback of existing unsupervised learning algorithms. Due to randomised initialisation, most existing methods require multiple training procedures to obtain the best performance.

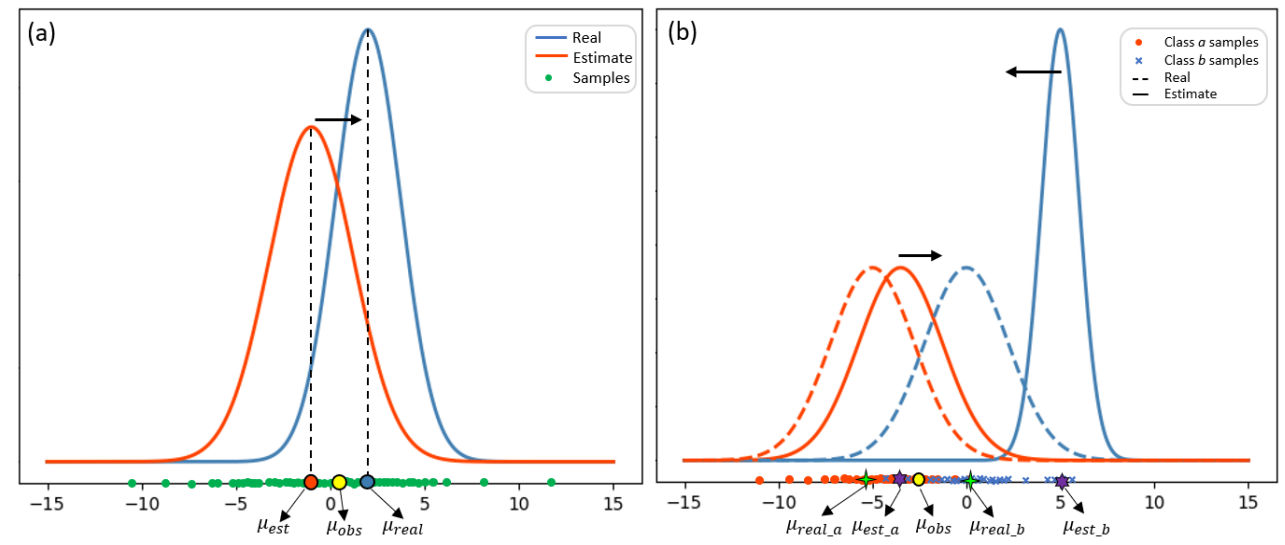

Fig. 2. Deviation of the estimated parameters: (a) normal distribution on single class samples (b) mixture model on multi class (number of class k=2) samples. Note that $\mu_{est}$ is the estimated mean value of the mixture model, $\mu_{obs}$ is the observed mean value of minibatch data $X$, and $\mu_{real}$ is the real (ideal) mean value of the mixture model.

Here, we propose a centralised constraint for the log-likelihood objective function to alleviate the degenerative issues of deep Gaussian networks. The objective function of the deep Gaussian network is calculated using the estimated parameters $\Theta$ and pseudo posterior $\gamma$. However, the variance in batch data makes it difficult to derive the real parameters $\mu_{real}$. To better demonstrate the idea of our proposed centralised constraint, two simplified examples are shown in Fig. 2. We first introduce a simplified scenario in Fig. 2 (a), which is a group of single-class samples following the Gaussian distribution. Given a batch of data $X$, let $\mu_{est}$ be the estimated mean value of the mixture model, $\mu_{obs}$ be the observed mean value of minibatch data $X$, and $\mu_{real}$ be the real (ideal) mean value of the mixture model. The centralised constraint will slightly drive $\mu_{est}$ close to the $\mu_{obs}$. Note that $\mu_{obs}$ does not equal to $\mu_{est}$ since it is the mean value of minibatch samples.

For multi-class samples, this centralised constraint can alleviate the negative effect of small inter-class variations (Fig. 2 (b)). Assume there are two classes $a$ and $b$, which denote $a'$ and $b'$ as the estimated classes. The model treats the majority samples of class $a$ and $b$ as the class $a'$, while some outliers of class $b$ are considered as $b'$. This could lead to poor segmentation results when performing existing methods on samples with small inter-class variations.

Therefore, a centralised constraint $\Delta$ is devised to let the estimated mean $\mu_{est}$ approximate $\mu_{obs}$ by considering the diversity of $X$

$$\Delta = \frac{|\mu_{est} - \bar{X}|}{{\sigma_X}^2}. \quad (5)$$

When dealing with hard samples with small inter-class variations, the observed variance is relatively small, resulting in a relatively large constraint value. This constraint will force the model to reallocate the estimated mean to approximate the observed mean; therefore, can reduce the degenerative issues. When dealing with "easy" samples (i.e., samples with large inter-class variations), the observed variance is high, leading to small constraints to the objective functions that can barely affect the parameter estimation.

With this centralised constraint $\Delta$, the objective function $\mathcal{L}_C$ for our DCGN can be expressed as

$$\mathcal{L}_C = \mathcal{L}(\Theta|\Theta^{(t)}) - \lambda \sum_{k=1}^{K}\sum_{c=1}^{C} \frac{|\mu_k{}^{(t)} - \overline{X_c}|}{\sigma_c{}^2}, \quad (6)$$

where $C$ is the dimension of the input samples (e.g., $C = 3$ for RGB images), $\sigma_c{}^2$ is the variance of minibatch samples on channel $c$, and $\overline{X_c}$ denotes the mean value of minibatch samples on channel $c$. With the proposed constraint, the objective function $\mathcal{L}_C$ would be penalised if the estimated $\mu_k$ is far away from the observed mean $\mu_{obs}$. As a result, outliers or hard samples would produce less interference to the objective function, hence, stabilising the training procedure, and in turn, improving the segmentation performance.

Assume the constraint weight as $\lambda$, by calculating partial derivatives over $\mu_k, \Sigma_k$ and $\alpha_k$ of Eq. (6), the centralised mixture parameters can be obtained via

$$\gamma_{i,k}^{(t+1)} = \emptyset(X, \omega^{(t)}) \quad (7)$$

$$\mu_k^{(t+1)} = \begin{cases} \dfrac{\left(\sum_{i=1}^{N}\gamma_{ik}^{(t+1)}X_i - \lambda\sum_{c=1}^{C}\dfrac{\Sigma_k{}^{(t)}}{\sigma_c{}^2}\right)}{\sum_{i=1}^{N}\gamma_{ik}^{(t+1)}}, & \mu_k \geq \overline{X_c} \\ \dfrac{\left(\sum_{i=1}^{N}\gamma_{ik}^{(t+1)}X_i + \lambda\sum_{c=1}^{C}\dfrac{\Sigma_k{}^{(t)}}{\sigma_c{}^2}\right)}{\sum_{i=1}^{N}\gamma_{ik}^{(t+1)}}, & \mu_k < \overline{X_c} \end{cases} \quad (8)$$

$$\alpha_k^{(t)} = \frac{\sum_{i=1}^{N}\gamma_{ik}^{(t+1)}}{N} \quad (9)$$

$$\Sigma_k^{(t)} = \frac{\sum_{i=1}^{N}\gamma_{ik}^{(t+1)}(X_i - \mu_k^{(t+1)})^T(X_i - \mu_k^{(t+1)})}{\sum_{i=1}^{N}\gamma_{ik}^{(t+1)}} \quad (10)$$

Note that in Eq. (10), the calculation of $\mu_k^{(t+1)}$ demands $\Sigma_k{}^{(t)}$; therefore, an initialisation of $\Sigma_k$ is required before the training process. A random initialisation from uniform distribution was used in this study.

**Algorithm 1**. Pseudo-code for training DCGN

**Input:** images $X \in \mathbb{R}^{W\times H\times 3}$
**Output:** trained network parameters $\omega$,
semantic prediction $\gamma$
1. randomly initialize $\Sigma_k^{(0)}$, network parameters $\omega^{(0)}$
2. **for** $t$ in iterations **do**
   $\gamma^{(t)} = \emptyset(X, \omega^{(t)}) \in \mathbb{R}^{W\times H\times K}$
   **update** $\mu_k^{(t+1)}, \alpha_k^{(i)}$ with $\gamma_k^{(t)}, \Sigma_k^{(t)}$
   **update** $\Sigma_k^{(t+1)}$ with $\gamma_k^{(t)}, \mu_k^{(t+1)}$
   Compute $\mathcal{L}_C$ through $\mu_k^{(t+1)}, \Sigma_k^{(t+1)}, \alpha_k^{(t+1)}$
   **update** $\omega$ by $\arg\min_{\omega}[-\mathcal{L}_C(\omega^{(t)})]$

The pseudo-code of the entire training procedure for DCGN is shown in Algorithm 1.

## C. Preprocessing

Each input image $X$ is pre-processed by the min-max normalisation through RGB channels, that is

$$X'_c = \frac{X_c - \max(X_c)}{\max(X_c) - \min(X_c)}, \quad (11)$$

where $X_c$ is the channel $c$ of the input image $X$.

# IV. Experiments

This section demonstrates all the experimental settings including datasets, evaluation metrics, implementation details and results. The efficiency of the proposed DCGN is assessed on a public dataset from the TCGA* repository (MoNuSeg†) and our in-house renal biopsy image (RBI) dataset.

## A. Datasets and Training Strategies

***MoNuSeg.*** MoNuSeg consists of 44 pathological tissue images with 28,846 manually annotated nuclear boundaries. These 1,000×1,000 images were extracted from the separate whole slide images (scanned at 40×) from the TCGA repository, representing 9 different organs from 44 individuals. The stromal and epithelial nuclei were manually labelled using Aperio ImageScope. Details of MoNuSeg are described in Table I. The various tissue sections greatly increase the richness and appearance variation of the dataset, which can provide a convincing assessment.

TABLE. I
COMPOSITION OF THE MONUSEG DATASET.

| Subset | Nuclei | Images | Anatomical Details |
|---|---|---|---|
| Training | 21623 | 30 | 6 breast, 6 liver, 6 kidney, 6 prostate, 2 bladder, 2 colon, 2 stomach |
| Testing | 7223 | 14 | 2 breast, 3 kidney, 2 prostate, 2 bladder, 1 colon, 2 lung, 2 brain |

***RBI.*** RBI includes more than 10,000 image patches extracted from 400 whole slide images with biopsy-proven results collected from the National Clinical Research Centre of Kidney Diseases, Jinling Hospital. All data were deidentified in accordance with the tenets of the Declaration of Helsinki [41]. Each image was resized to a unified size of 512×512. We randomly selected 577 images for training and 20 images for validation (the glomerular structures were annotated by experienced pathologists with 20 years of experience). Note that the training set and validation set were selected from different whole slide images.

***Training Strategies.*** Parameters of the encoder are initialised with ImageNet pre-trained weights to provide strong feature extraction capabilities, while that of the decoder are initialised using He-normal initialisation. Randomised hue transformation (delta=0.12), randomised saturation (saturation factor ranges from 0.5 to 1.5), randomised flip-up/down, and randomised flip-left/right were implemented to augment the dataset before training. All of the models were trained on an NVIDIA RTX 3090 GPU for 200 epochs, with an initial learning rate of $5e^{-5}$ and a decay of 0.98 per epoch.

## B. Experimental Details

***Comparisons***. To evaluate the effectiveness of DCGN, we compared it with several deep learning based and conventional unsupervised segmentation methods, including minibatch K-Means (denote as mKMeans), GMM, IIC [31], Double DIP [34], DCAGMM (deep clustering via adaptive GMM modelling) [42], DIC (deep image clustering) [43], Kim's work [29], Kanezaki's work [14] and DCGMM [37]. It is of note that we reproduce and modify the DCAGMM by

* The Cancer Genome Atlas (TCGA), [Online]. Available at: http://cancergenome.nih.gov/ (Accessed in August. 2021)
† The MoNuSeg public dataset [Online]. Available at https://monuseg.grand-challenge.org/Data/ (Accessed in July 2021)

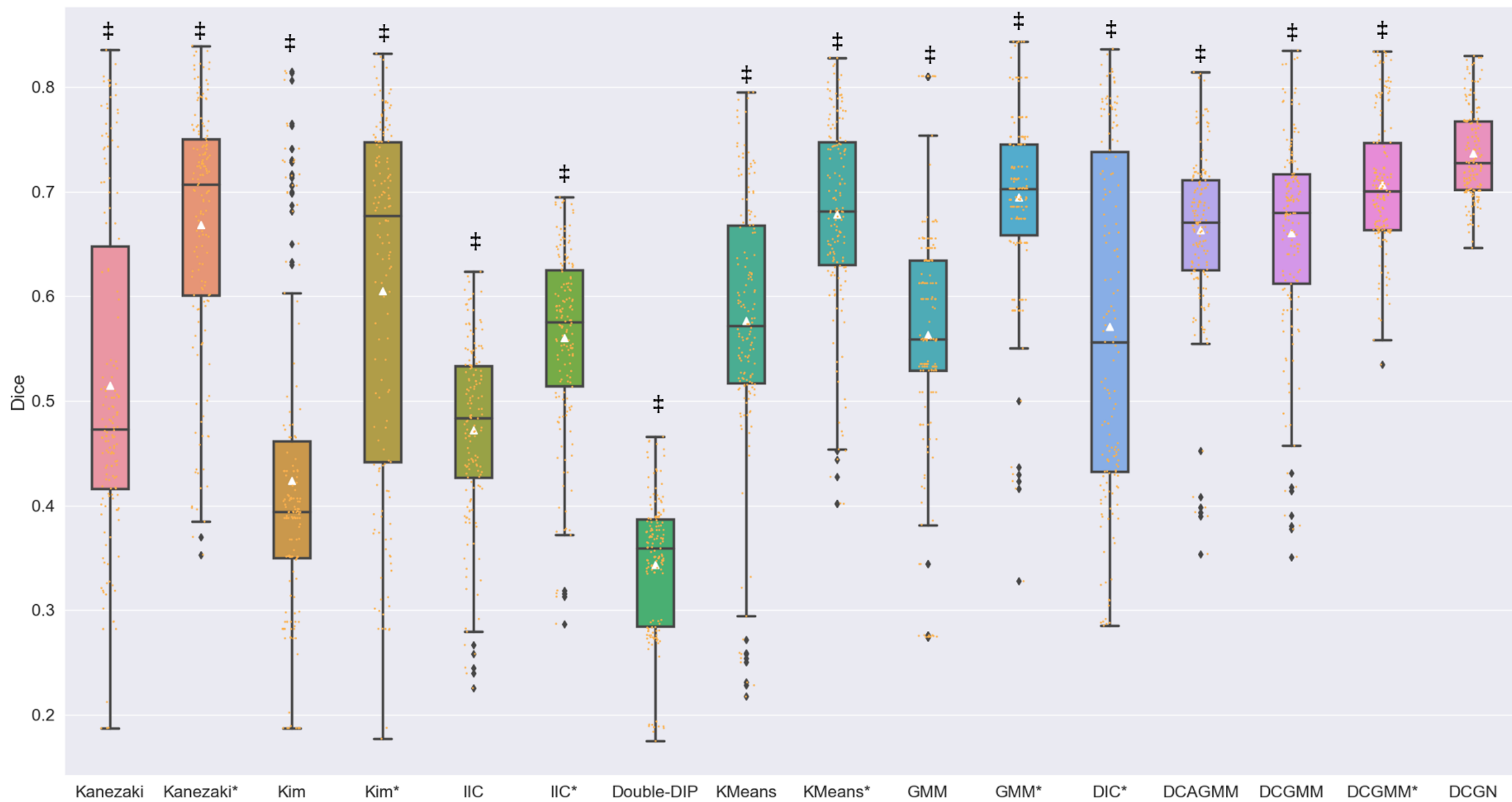

Fig. 3. Box plot of the Dice score during repeated experimental studies, where * denotes the model with redundant class (the number of pre-defined classes k=3 for cell segmentation), ‡ indicates highly significant differences results (Wilcoxon signed-rank test with P<0.001) compared with DCGN, the black dots refer to outliers and white triangles indicate mean values, the small orange dots refer to samples.

TABLE. II
PERFORMANCE OF THE CELL SEGMENTATION (MONUSEG DATASET).

| Methods | Precision | Recall | Dice | AJI |
|---|---|---|---|---|
| mKMeans* | 0.657±0.175(0.679)$^{\star}$ | 0.792±0.174(0.773)$^{\ddagger}$ | 0.678±0.094(0.682)$^{\ddagger}$ | 0.305±0.140(0.338)$^{\ddagger}$ |
| GMM* | 0.631±0.150(0.664)$^{\ddagger}$ | 0.822±0.109(0.819) | 0.695±0.085(0.717)$^{\ddagger}$ | 0.290±0.151(0.319)$^{\ddagger}$ |
| IIC* | 0.467±0.092(0.516)$^{\ddagger}$ | 0.725±0.121(0.796)$^{\ddagger}$ | 0.560±0.087(0.618)$^{\ddagger}$ | 0.056±0.030(0.072)$^{\ddagger}$ |
| Kim et al.* | 0.575±0.249(0.698)$^{\ddagger}$ | 0.824±0.189(0.772) | 0.606±0.171(0.694)$^{\ddagger}$ | 0.220±0.176(0.323)$^{\ddagger}$ |
| Double DIP | 0.221±0.051(0.221)$^{\ddagger}$ | 0.820±0.109(0.851) | 0.344±0.067(0.350)$^{\ddagger}$ | 0.013±0.006(0.013)$^{\ddagger}$ |
| Kanezaki et al.* | 0.629±0.195(0.725)$^{\ddagger}$ | 0.822±0.162(0.783) | 0.669±0.119(0.727)$^{\ddagger}$ | 0.260±0.166(0.351)$^{\ddagger}$ |
| DCGMM* | **0.693±0.135(0.698)**$^{\star}$ | 0.786±0.171(0.801)$^{\ddagger}$ | 0.707±0.064(0.719)$^{\ddagger}$ | 0.314±0.124(0.345)$^{\ddagger}$ |
| DIC* | 0.511±0.249(0.595)$^{\ddagger}$ | **0.848±0.170(0.832)**$^{\star}$ | 0.571±0.165(0.644)$^{\ddagger}$ | 0.147±0.169(0.193)$^{\ddagger}$ |
| DCAGMM | 0.619±0.137(0.691)$^{\ddagger}$ | 0.767±0.131(0.763)$^{\ddagger}$ | 0.664±0.079(0.706)$^{\ddagger}$ | 0.300±0.126(0.365)$^{\ddagger}$ |
| DCGN | 0.685±0.113(0.716) | 0.834±0.115(0.808) | **0.737±0.043(0.743)** | **0.352±0.113(0.379)** |
| U-Net$^{\dagger}$ | 0.695±0.095(0.740) | 0.849±0.083(0.848) | 0.755±0.045(0.782) | 0.370±0.093(0.436)$^{\star}$ |

* denotes redundant class (k=3) and † refers to a fully supervised learning baseline using modified U-Net. The bold values refer to the best average performance among unsupervised methods (without considering supervised U-Net). $^{\star}$(‡) indicates significant differences (highly significant differences) results compared with DCGN, with Wilcoxon signed-rank test P<0.05 (P<0.001). The results are shown as "mean± standard deviation (upper-bound results)".

adopting its distance-based constraints in the original DCGMM (it was initially designed for image classification). Open-source implementations of the comparison methods used in this study can be obtained on Github. The network structure of the DCGMM was modified to match our DCGN for a fair comparison. In addition to unsupervised methods, we also implemented a fully supervised U-Net on cell segmentation task for better comparison. The implemented U-Net was modified by adding batch normalization layers and dropout layers compared to the original vanilla U-Net [44].

***Cell Segmentation on MoNuSeg.*** For many existing unsupervised learning approaches, the performance of segmentation suffers from random initialisation. In this study, repeated experiments were conducted to explore the stability and reproducibility of the performance of all comparison algorithms. All these approaches were trained for 150 epochs each time and repeated 10 times without changing any parameters or training samples. The upper bound performance is defined as the best results among 10 repeated experiments. Although cell segmentation is a binary task, all the compared studies were assessed using different numbers of classes (k=2 or 3) to show their upper-bound performance. In addition, a fully supervised U-Net is trained as the baseline of supervised learning.

***Glomeruli Decomposition on RBI.*** In addition to assessing the effectiveness of the binary segmentation, a glomeruli decomposition task is carried out. The glomerular structures were divided into three parts (k=3), including (1) mesangial matrix and basement membrane, (2) intra-glomerular cells (mesangial, endothelial and podocytes) and macula densa, and (3) other regions such as glomerular capillaries, bowman's space, exudate, etc. It is of note that Double DIP was not assessed since it was designed for binary segmentation only.

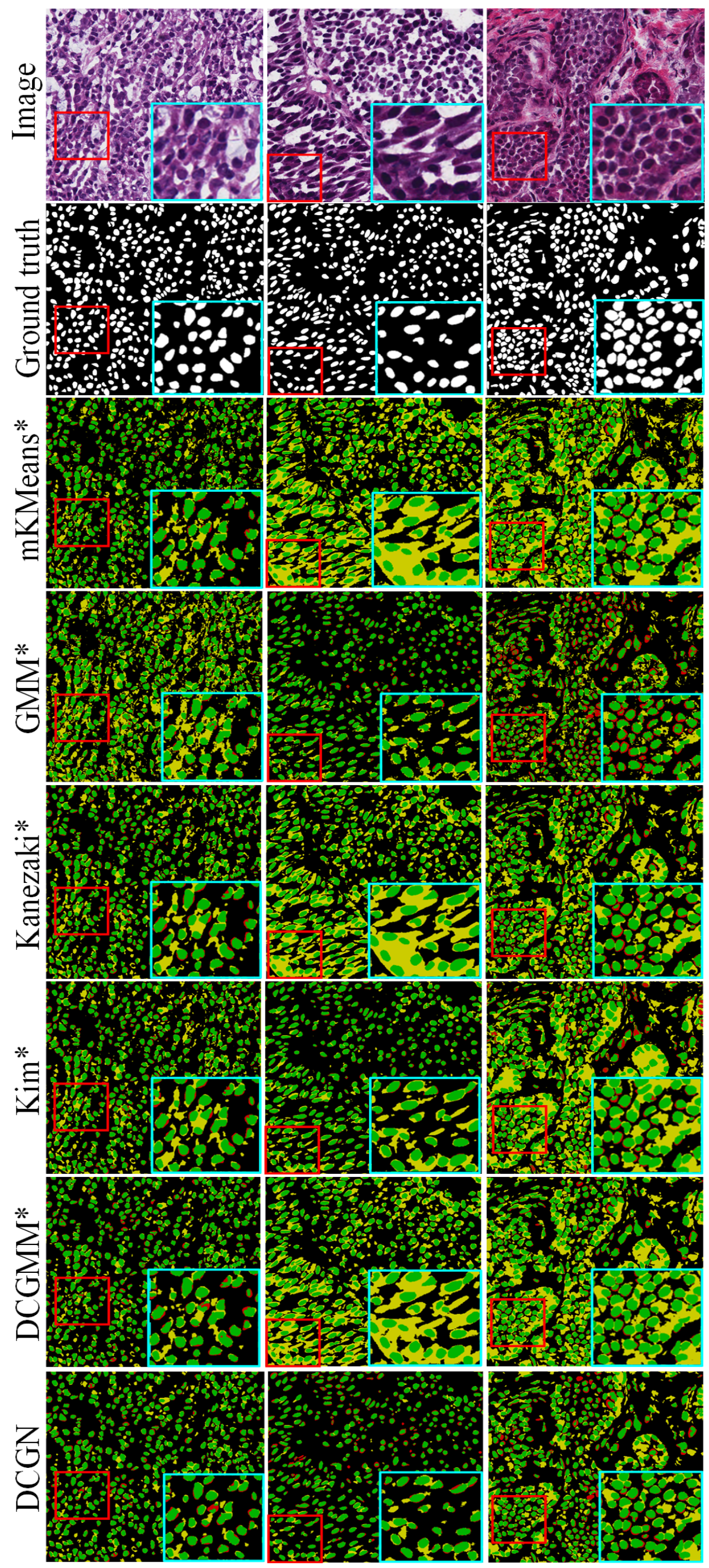

Fig. 4. Comparison of unsupervised cell segmentation results, where * denotes models with redundant class (k=3). Green, yellow, and red colours refer to the true positive, the false positive and the false negative predictions, respectively. The red and cyan boxes highlight the region of interests before and after zoom-in.

***Degeneration Assessment.*** To explore the degenerative issues, we analysed 140 predictions on the MoNuSeg datasets and 100 predictions on the RBI datasets, based on the following criteria:

(1) All these predictions are acquired from repeated experiments (10 times for MoNuSeg and 5 times for RBI).

(2) Collapse is assessed on both ***MoNuSeg*** and ***RBI*** datasets, which is defined as a certain class dominating the major region (here we set 97% as the threshold) of an image.

(4) Redundant class is assessed on the ***MoNuSeg*** dataset, which is identified when the segmentation performance can be improved by adding an extra class without semantic meanings.

(5) Empty class is assessed on the ***RBI*** dataset and refers to missing a certain class or with an extremely low ratio (here we set <1%) in the prediction.

(6) Instability is assessed on both ***MoNuSeg*** and ***RBI*** datasets and is considered when the standard deviation of the average performance among repeated experiments is larger than 8%.

***Evaluation Metrics***. In addition to the commonly used Dice coefficient score, pixel-wise precision and recall were also reported. To statistically evaluate the performance, Wilcoxon signed-rank test was adopted between the evaluation results derived using DCGN and other comparison methods, with P<0.05 (or P<0.001) indicating significant (or highly significant) differences between the two paired methods. The Aggregated Jaccard Index (AJI) was applied to the MoNuSeg dataset to verify the instance-level segmentation performance, that is

$$\mathrm{AJI} = \sum_{i=1}^{N} \frac{G_i \cap P_i}{G_i \cup P_i + \varepsilon}, \tag{12}$$

where $i$ indicates the number of cells, $\varepsilon$ is the smooth parameter, $G_i$ and $P_i$ refer to the ground truth and prediction of the $i$-th cell. In glomeruli segmentation, we applied normalised mutual information (NMI) to assess the mutual dependence between two samples, which is given by

$$\mathrm{NMI}(Y, C) = \frac{2I(Y; C)}{[H(Y) + H(C)]}, \tag{13}$$

where $Y$ refers to the ground truth labels and $C$ denotes the prediction, and $I$ is the mutual information of $Y$ and $C$, $H(.)$ is the entropy. It is of note that all the ground truth labels were only used during the evaluation that had not been revealed in the training process.

## C. Experimental Results

***Unsupervised Cell Segmentation on MoNuSeg.*** The performance of repeated experiments is presented in Table II, shown as mean ± standard deviation (with the upper-bound results of each method shown in brackets). It shows that some unsupervised approaches initially developed for natural images could not perform well on pathological images, indicating a significantly lower average Dice (relatively 3-39% lower) compared to the proposed DCGN (Fig. 3 and Table II). For instance, double DIP [34] failed to perform cell segmentation with only a 0.344 average Dice score. Interestingly, conventional GMM (k=3) achieved good performance with a 0.695 average Dice score, which is similar compared to that of the DCGMM (0.707).

To provide statistical assessments, Wilcoxon signed-rank test was performed between the evaluation results of 10 repeated experiments. Considering the upper bound of the segmentation performance (shown in Table II), the proposed DCGN achieved the best Dice score (0.743) among unsupervised learning approaches, followed by Kanezaki's (0.727) and DCGMM (0.719). Moreover, DCGN achieved the best AJI score (0.379) among all the unsupervised learning approaches.

In addition, DCGN achieved a significantly better Dice coefficient score and AJI score compared to other unsupervised segmentation approaches ($P<0.001$). Interestingly, there were no significant differences ($P>0.05$) found for Precision, Recall and Dice scores using our DCGN compared to the fully supervised U-Net based method (Table II). Although the DCGMM achieved better Precision compared to our DCGN ($P=0.036$), its Recall, Dice and AJI score are significantly lower than the proposed DCGN ($P<0.001$). DIC has the highest Recall, but relatively low Precision indicating lots of false-positive predictions. Double DIP achieved a high recall as well but the lowest precision score and therefore a very low Dice score. To better demonstrate the performance of the competitive approaches (Dice>0.65), three images were randomly selected from the test set to visualise the upper-bound segmentation performance (Fig. 4). It is of note that in Fig. 4, predictions of the redundant class have been removed (some methods achieved upper-bound performance by adding a redundant class (i.e., k=3)).

TABLE. III
PERFORMANCE OF THE GLOMERULUS SEGMENTATION (RBI DATASET).

| Methods | NMI | Dice |
|---|---|---|
| mKMeans | 0.200±0.040(0.200) ‡ | 0.555±0.037(0.559) ‡ |
| GMM | 0.328±0.051(0.328) ‡ | 0.640±0.061(0.640) ‡ |
| DCGMM | 0.207±0.042(0.229) ‡ | 0.567±0.047(0.579) ‡ |
| Kanezaki | 0.186±0.095(0.207) ‡ | 0.502±0.109(0.537) ‡ |
| IIC | 0.090±0.068(0.124) ‡ | 0.501±0.054(0.534) ‡ |
| Kim | 0.187±0.089(0.195) ‡ | 0.500±0.106(0.511) ‡ |
| DIC | 0.192±0.117(0.234) ‡ | 0.516±0.127(0.558) ‡ |
| DCAGMM | 0.207±0.041(0.212) ‡ | 0.578±0.048(0.582) ‡ |
| DCGN | **0.377±0.053(0.384)** | **0.735±0.050(0.746)** |

‡ indicates highly significant differences results (Wilcoxon signed-rank test $P<0.001$) compared with our DCGN. The bold values refer to the best performance among comparison methods. The results are shown as "mean± standard deviation (upper-bound results)".

***Unsupervised Glomeruli Decomposition on RBI.*** The average performance of our comparison study on RBI is summarised in Table III, Fig. 5, assessed by NMI and Dice coefficient score. All comparison studies were performed with k=3 to segment three semantic labels (the definition of semantic labels is described in Section IV B).

As Table III shows, the proposed DCGN achieved significantly better results ($P<0.001$) compared to state-of-

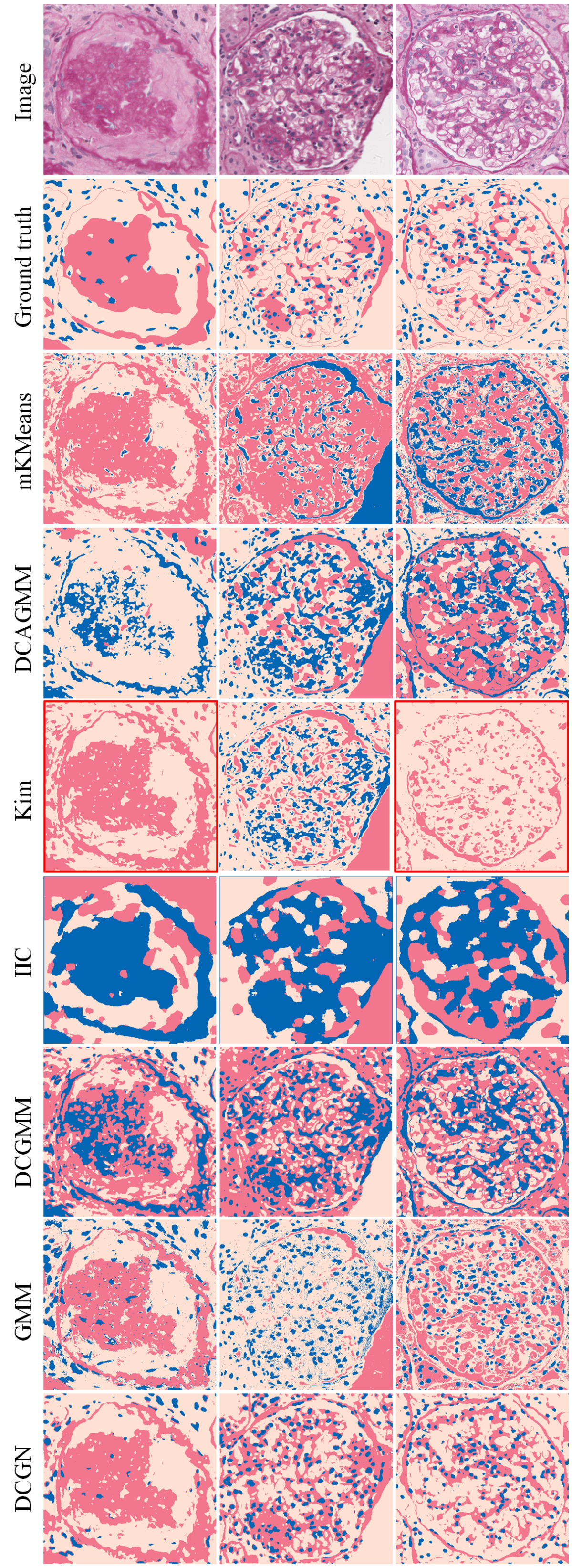

Fig. 5. Comparison of unsupervised glomeruli segmentation results. Empty class issues are highlighted by red bounding boxes. The red, blue, and almond colours in the ground truth refer to (1) mesangial matrix and basement membrane, (2) intra-glomerular and macula densa cells, (3) other regions such as glomerular capillaries, respectively.

TABLE. IV

DEGENERATION ASSESSMENT (MONUSEG AND RBI DATASET)

| | Collapse | Empty Class | Stability |
|---|---|---|---|
| GMM | 0/240 | 0/100 | × |
| Kanezaki | 7/240 | 23/100 | × |
| Kim | 17/240 | 28/100 | × |
| mKMeans | 0/240 | 0/100 | × |
| IIC | 0/240 | 0/100 | × |
| DIC | 0/240 | 25/100 | × |
| DCAGMM | 0/240 | 1/100 | √ |
| DCGMM | 0/240 | 2/100 | √ |
| DCGN | 0/240 | 0/100 | √ |

The red blocks indicate the occurrence of degenerative issues.

the-art methods on glomeruli composition, with an average of 0.735 Dice score and 0.377 NMI, followed by GMM (an average of 0.640 Dice and 0.328 NMI) and DCAGMM (an average of 0.578 Dice and 0.207 NMI).

***Degeneration Assessment***. The results of the degeneration assessment are shown in Table. IV. It is of note that Double DIP was not assessed due to its relatively weak performance. As Table. IV shows, the methods proposed by Kanezaki and Kim heavily suffered from all degenerative issues. Similarly, the empty class is prone to occur in DIC. DCGMM and DCAGMM occasionally encountered the empty class issue and GMM presented instability during repeated experiments. Both mKMeans and IIC witnessed instability in the MoNuSeg dataset.

## V. Discussion

In this study, we have developed a novel unsupervised segmentation method combining deep neural networks with a constrained GMM. This approach has been comprehensively evaluated on pathological images using both a public MoNuSeg dataset and an in-house RBI dataset. We have achieved significantly better results compared to previously published unsupervised segmentation methods with clear evidence of mitigating degenerative issues that are currently challenging for pathological tissue image delineation. Besides, our proposed method has also achieved comparable results with some widely used semi-supervised and fully supervised learning methods.

***Performance Analysis.*** Comprehensive comparison results in Tables II and III and Figs. 4 and 5 have demonstrated the superior segmentation capability of the proposed DCGN. Compared to existing unsupervised segmentation methods, our DCGN is robust to small inter-class variations. For instance, as Fig. 5 (second column) shows, all the unsupervised methods except DCGN have regarded white regions as a single class while ignoring the exudation/stroma regions (light pink regions in the raw images).

Interestingly, conventional methods such as mKMeans and GMM have shown their effectiveness in tissue segmentation. In particular, GMM has obtained better performance than mKMeans for tissue segmentation with slightly worse stability. It achieved better performance than mKMeans in kidney tissue segmentation, with a 0.08 higher average Dice score and 0.12 higher NMI score, respectively. Methods proposed by Kanezaki et al. and Kim et al. have produced reasonable results on cell segmentation but have suffered heavily from collapse and empty class issues (large variances in Fig. 3 and many failed cases summarised in Table IV). We observed poor segmentation for these two methods when dealing with kidney tissue segmentation (see Table. III and Fig. 5). DIC have presented a high recall score with a low precision score in cell segmentation and poor results in glomeruli segmentation. Double DIP has derived similar coarse predictions (high recall but low precision scores) as DIC for cell segmentation, indicating its incompatibility for tissue segmentation, although the method could be more adaptive for natural image segmentation. The coarse predictions given by IIC have indicated its inapplicability to pathological images. Although DCGMM has presented comparable performance to our DCGN on cell segmentation, it has achieved significantly lower segmentation accuracy on kidney tissue segmentation and has issues with generating empty classes. Moreover, DCGMM has presented poor performance when dealing with samples with small inter-class variations (poor cell segmentation results from dark background areas as shown in Fig. 4 middle column). Similar to DCGMM, DCAGMM presented comparable results. However, its normalized distance constraint (which aims to increase the distance between Gaussian centres) makes it hard to segment classes with high intra-class variations.

***Comparing with Fully Supervised Segmentation Methods.*** One of the major concerns of unsupervised segmentation is how it performs compared with fully supervised segmentation algorithms. In addition to the U-Net baseline given in Table III, we compared the proposed DCGN with previously published supervised studies (Table V). It is of note that all comparisons were performed on the same test data of the MoNuSeg dataset. As Table V shows, the proposed DCGN has achieved a comparable average Dice coefficient score compared with the fully supervised U-Net based method (no significant differences were found in the Precision, Recall and Dice score). DCGN has obtained significantly better performance compared to other unsupervised segmentation methods (Tables II and III), it, however, has presented a lower AJI score compared to fully supervised and semi-supervised segmentation methods (Table V). This is mainly because of the adhesion of adjacent cells, which could be better addressed using supervised or semi-supervised methods.

TABLE. V

PERFORMANCE OF SUPERVISED METHODS (MONUSEG DATASET).

| Methods | Avg F1 (Dice) | Avg AJI |
|---|---|---|
| DCGN | 0.7432 | 0.3790 |
| U-Net | 0.7582 | 0.4357 |
| Mask RCNN [45] * | 0.7991 | 0.5128 |
| Dual U-Net [46] * | 0.7913 | 0.5899 |
| Tian et al. [47] †, * | 0.7638 | 0.4927 |
| Qu et al. [48] †, * | 0.7566 | 0.5160 |
| CNN [49] * | 0.7623 | 0.5083 |

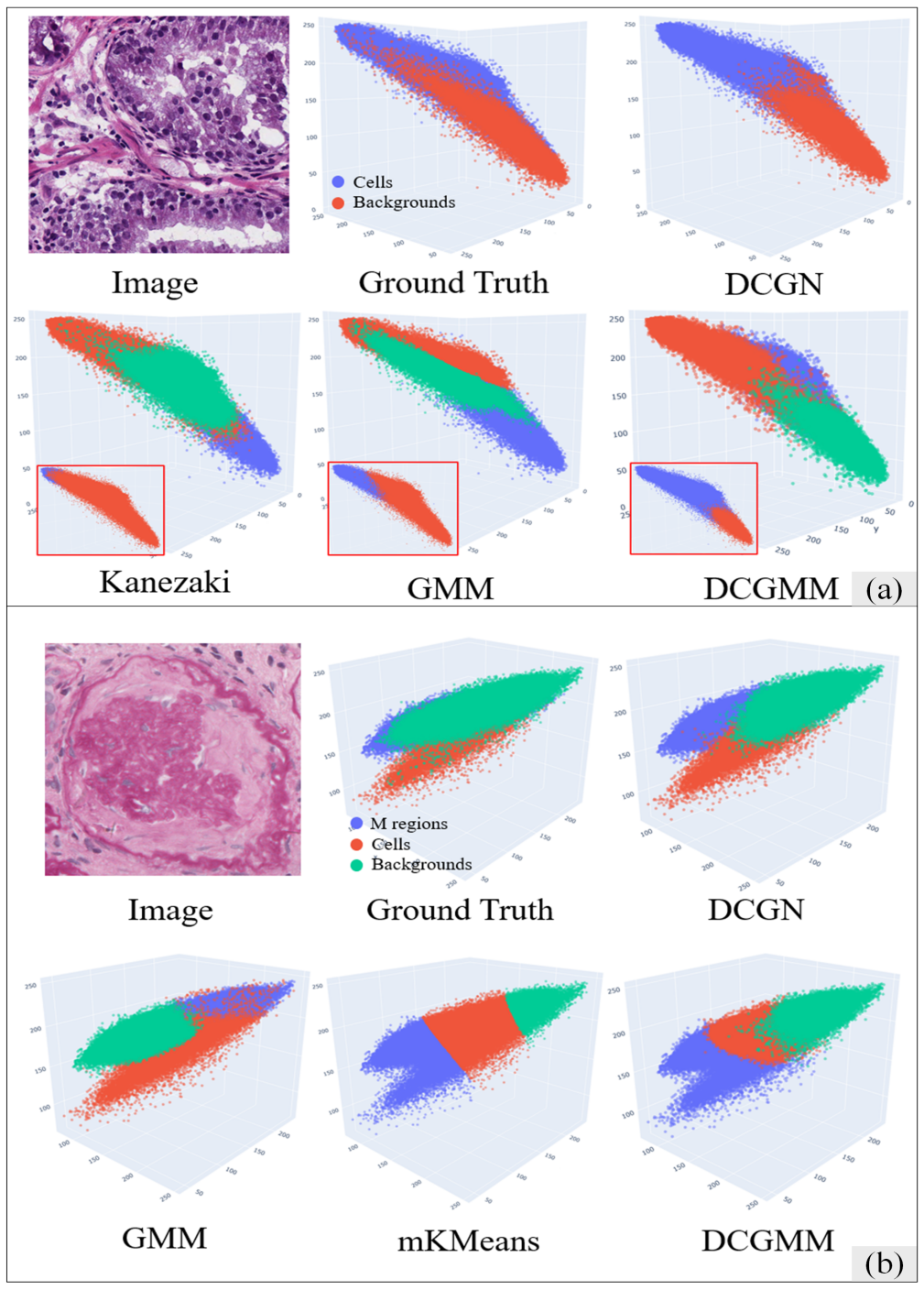

Fig. 6. Class intensity maps of the top 4 methods for (a) cell segmentation and (b) renal tissue segmentation. The three axes refer to the R, G, and B intensities, and different colours denote different classes. The vignette in red boxes in (a) indicate class intensity maps without redundant class.

* indicates patch based training progress, † refers to semi-supervised learning approaches.

Both semi-supervised approaches proposed by Tian et al. [47] and Qu et al. [48] have taken prior knowledge of cell central points into account, leading to competitive AJIs of 0.4927 and 0.5160. In addition, significant improvement in average AJI has been observed using patch-based methods (denote with * in Table V) compared to the raw image-based learning strategy. This has indicated the importance of the patch learning strategy in the tissue segmentation task (here patch-based methods refer to extracting small patches from original raw images in both the training and testing process). Overall, it can be difficult for unsupervised segmentation approaches to produce precise pixel-level predictions, especially for dense and small objects.

***Distinguishing Samples with Small Inter-Class Variations.*** The capability of distinguishing small inter-class variation samples determines the accuracy of the subtle tissue segmentation. We have explored this capability by plotting the class intensity map of the top 4 methods in cell and kidney tissue segmentation, respectively. As Fig. 6 (a) shows, most unsupervised methods have not been able to clearly segment the background samples and require a redundant class for those hard samples, while DCGN can effectively distinguish background samples and foreground samples without adding a redundant class. As shown in Fig. 6 (b), mKMeans method presented hard boundaries due to the Euclidean distance measurement, while other methods have produced smoother boundaries. DCGN has presented the most similar class intensity maps compared to the ones generated from the ground truth, indicating the effectiveness of the proposed centralised function.

***Redundant Class.*** Experimental results have indicated that most unsupervised segmentation methods have suffered from the redundant class issue. As Fig. 3 shows, most of the compared methods have obtained a significant performance improvement for the binary segmentation task when changing the number of classes from 2 to 3. The reason behind this is that these models can be struggling to distinguish samples with small inter-class variations. While the pre-defined number of classes cannot well accommodate all samples, unstable performance can be observed since the hard samples can be assigned with different labels at different repeated experiments. For example, white background pixels may be assigned as background samples in the first round of training while assigned as the foreground samples in another round. Therefore, these unsupervised methods require a redundant class to accommodate these hard samples. However, our DCGN has the capability for accurate tissue segmentation without using an additional redundant class that is more efficient and effective.

***Stability.*** As shown in Fig. 3 and Table V, IIC, mKMeans, DCGMM and DCGN have presented good stability in repeated experiments. Similar to the conventional GMM that has suffered from instability, the performance of Kim's and Kanezaki's methods has also presented dramatic fluctuation with large variances. In addition, the stability of previous methods has been enhanced by introducing a redundant class to accommodate hard samples. However, even though IIC, mKMeans and DCGMM have presented good stability, their segmentation performance has been significantly lower than our DCGN.

***Reproducibility and Empty Class Issues.*** Methods that cannot be trained on large-scale studies are more likely to result in poor reproducibility. For instance, conventional GMM without minibatch learning can only be performed on a small number of images. This leads to limited information when developing generalised segmentation models. Moreover, some methods (e.g., Kim's and Kanezaki's methods) can only produce a single image during the training process, leading to low reproducibility of repeated experiments (i.e., obtaining the same semantic labels for the same samples).

The empty class problem is another issue that has hindered the deployment of unsupervised segmentation. For instance, Kim's, Kanezaki's and DCGMM methods have encountered empty class issues during the evaluation. This is caused by the incapability of separating hard samples (i.e., delineation of pixels with similar intensities but different categories). In contrast, the proposed DCGN can effectively avoid the empty class issue and achieve higher reproducibility in large-scale training.

**Ablation Studies of Penalty Weights.** The influence of the proposed centralised constraint is explored by setting different weights $\lambda$ in Eq. (6). The results of 10 repeated experiments (for each $\lambda$) are shown in Table V.

TABLE. VI

Ablation studies of constrained weights

| $\lambda$ | Dice | Avg Epochs |
|---|---|---|
| 0.05 | 0.637±0.076 (0.740) | 37 |
| 0.005 | 0.737±0.043 (0.743) | 62 |
| 0.0005 | 0.734±0.005 (0.745) | 89 |

"Avg Epochs" indicates the average number of epochs for convergence.

It can be observed that the upper bound performance of models with different $\lambda$ remains similar, with 0.740, 0.743, 0.745 of $\lambda$=0.05, $\lambda$=0.005 and $\lambda$=0.0005, respectively. However, the standard deviation of the Dice score exerts significant differences. As Table VI shows, a large weight for the centralised constraint leads to faster convergence while also leading to an unstable training procedure (which may be attributed to the local optimum trapping of the module). A smaller weight requires more training epochs for convergence but has more stable training processes.

***Capacity on whole slide images.*** It remains unclear how DCGN performs on whole slide images when predictions are made across patches (tiles). Here we tested the cell segmentation module (two classes) on a renal whole slide image. It demonstrated that our method could achieve promising performance when handling renal images with homogenous features. However, false-positive samples could be observed in some vessel regions, indicating potential research directions (e.g., enhancing the utilization of textural features) to improve the module capacity.

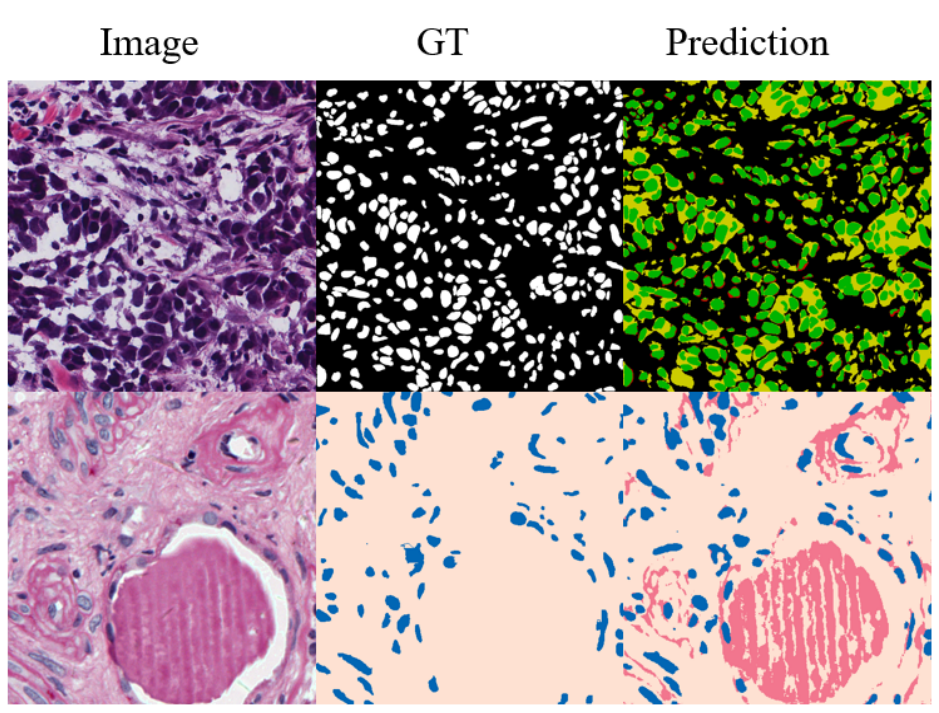

Fig. 7. Weak predictions of cells and glomerular structures.

***Limitations.*** The essence of unsupervised learning is to allocate the same label to samples of the same class. However, it is almost impossible to acquire precise segmentation predictions without any prior knowledge or annotation. Compared with the existing studies [50, 51] of pathological image segmentation, the proposed method may not able to produce satisfactory instance segmentation results (cells are prone to adhesion), which may limit its clinical application when a single-cell analysis is necessary. Most of the unsupervised learning methods are performed based on pixel intensities without considering textual features. Although combining deep neural networks with clustering or mixture models can enhance the utilization of textual features, it still relies on pixel intensity-based objective functions to some extent. The weak predictions can be observed in the segmentation of cells (first row in Fig. 7.) and glomerular structures (second row in Fig. 7.). This is mainly because of the conflict between the hypothesized Gaussian and real data distributions. Although the proposed DCGN may not be able to produce satisfactory predictions when handling complex images with too many categories or images with many "outliers", the DCGN has shown merits in upstream (general tasks such as foreground/background segmentation) tasks. More importantly, the proposed constraint can help the module to build better classification boundaries for classes with small inter-class variations which is a major technical contribution of our method; however, our method can alleviate the false predictions but not completely remove them.

**How does DCGN alleviate degenerative issues?** In order to give readers more intuition about how our method addresses the degenerative issues, we designed some schematic illustrations using simplified examples in 2D space (because real 3D cluster are intricate to demonstrate and comprehend).

First, the missing class issue usually occurs when the module fails to address the outliers, e.g., the module takes the outliers as a unique class while combing certain categories (blue and red dots) into a single class (as shown in Fig.8(a)). This kind of issue is more likely to occur in iterative methods that rely on pseudo labels, while it is also occasionally witnessed in existing deep Gaussian networks. The proposed centralised constraint will force the mixture module to be closer to the centroid of the data samples, thus preventing the occurrence of the missing class issue.

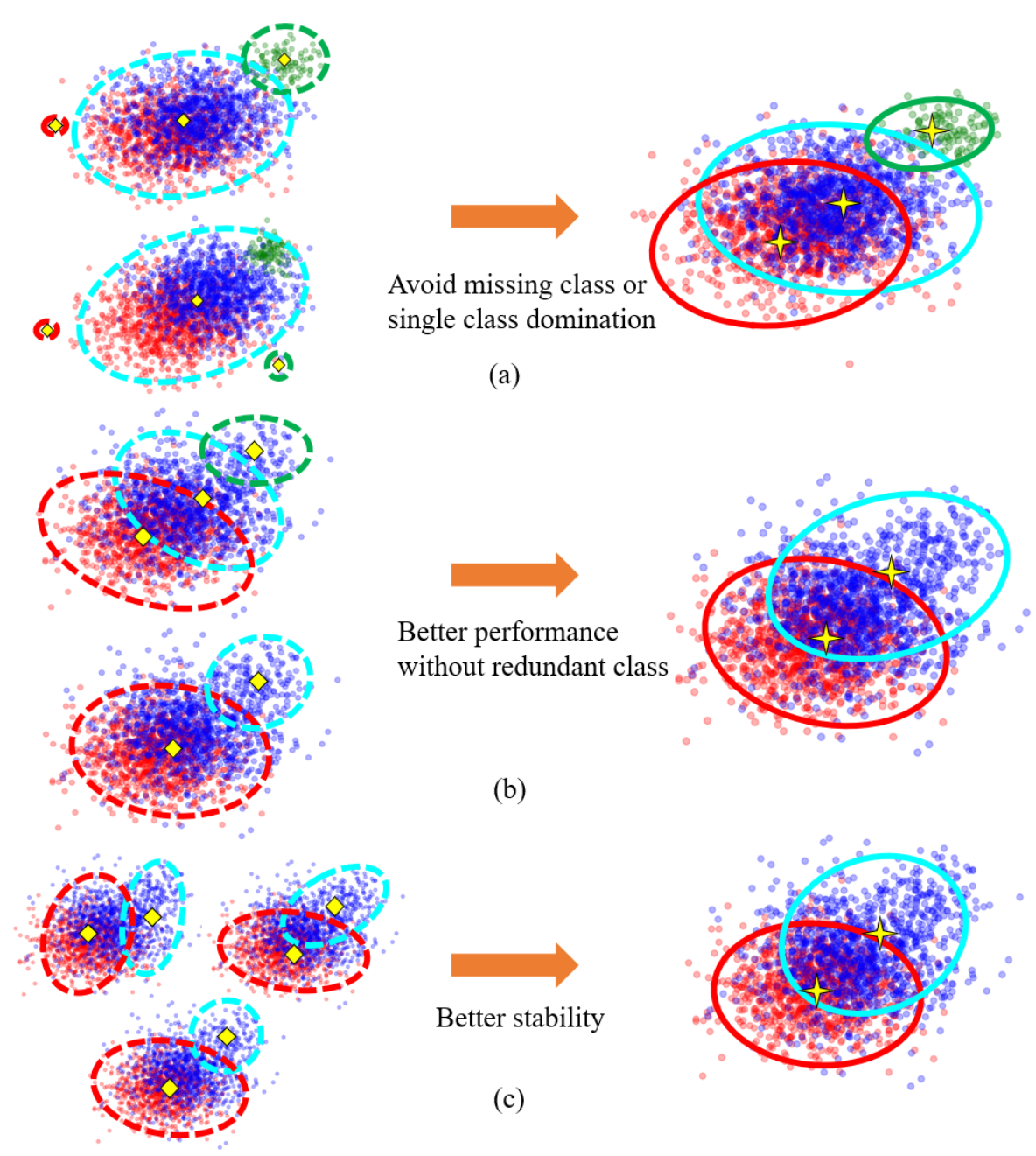

Fig. 8. Simplified examples to illustrate how the proposed centralised constraint addresses the (a) missing class and single class domination (collapse); (b) redundant class and (c) instability problems. Predictions given by methods without centralised constraint are noted with dotted circles (left column). Class centroids are shown as yellow diamonds (class centroid given by methods without centralised constraint) and yellow stars (class centroid given by the proposed method).

Second, the redundant class issue is usually artificial, as to improve the performance of most unsupervised methods. Due to the discrete distribution of a certain class (e.g., background regions that contain stroma and white non-tissue areas), some methods may need an additional class to 'collect' certain samples (shown as the blue samples within the green dotted circles in Fig.8 (b)). The redundant class can be simply avoided by setting an appropriate number of classes, however, modules without centralised constraints cannot achieve good performance (as shown in Fig. 3).

Moreover, the instability (low reproducibility) occurs because of the random initialisation. The proposed centralised constraint can alleviate the randomness caused by initialisation since it forces the module to learn parameters that approximate the data centroid (the proposed method achieves the lowest variance of evaluation metrics as shown in Table II.).

**Suggested criteria and Future Directions.** Based on the findings of our study, we emphasize these in-depth evaluation criteria for unsupervised segmentation approaches: 1) Repeated experiments should be conducted to present the stability and reproducibility of the method and 2) The degenerative issues should be discussed in detail to check the robustness of the method.

Here we also provide some potential research directions for unsupervised segmentation. The proposed DCGN can address essential segmentation tasks in pathological images. However, there remains further exploration on how it performs on other image modalities, e.g., segmenting the tumour from brain magnetic resonance scans [52, 53] or segmenting organs from computerised tomography images [54]. In addition, the uncertainty estimation of the semantic predictions for unsupervised segmentation should be explored. By using those 'confident' predictions, a self-supervised paradigm may be integrated with unsupervised learning to achieve superior performance. Moreover, methods that can cope with images with many classes still need to be developed, since most unsupervised segmentation approaches can only deal with relatively simple semantic predictions (e.g., learning by imitation to address the unseen classes [55]). Last but not least, a robust model that can better address the "outliers" should be developed.

## VI. Conclusion

Tissue segmentation is an essential step of computational pathology; however, most existing methods demand a large number of manual annotations. This study demonstrates an effective unsupervised tissue segmentation using the developed, innovative DCGN method. The proposed DCGN method can accurately segment tissue structures without using any manual annotations or prior knowledge. This could potentially reduce the annotation costs in computational pathology dramatically.

## Acknowledgements

This study was supported in part by the ERC IMI (101005122), the H2020 (952172), the MRC (MC/PC/21013), the Royal Society (IEC\NSFC\t211235), the NVIDIA Academic Hardware Grant Program, and the UKRI Future Leaders Fellowship (MR/V023799/1). We acknowledge Michael Yeung from Cambridge University for proofreading the manuscript.

## References

[1] P. U. Adiga and B. Chaudhuri, "An efficient method based on watershed and rule-based merging for segmentation of 3-D histo-pathological images," *Pattern recognition,* vol. 34, no. 7, pp. 1449-1458, 2001.

[2] S. Wienert *et al.*, "Detection and segmentation of cell nuclei in virtual microscopy images: a minimum-model approach," *Scientific reports,* vol. 2, no. 1, pp. 1-7, 2012.

[3] I. Nogues *et al.*, "Automatic lymph node cluster segmentation using holistically-nested neural networks and structured optimisation in CT images," in *International Conference on Medical Image Computing and Computer-Assisted Intervention*, 2016: Springer, pp. 388-397.

[4] S. Ragothaman, S. Narasimhan, M. G. Basavaraj, and R. Dewar, "Unsupervised segmentation of cervical cell images using gaussian mixture model," in *Proceedings of the IEEE conference on computer vision and pattern recognition workshops*, 2016, pp. 70-75.

[5] Y. Zhang, M. Brady, and S. Smith, "Segmentation of brain MR images through a hidden Markov random field model and the expectation-maximization algorithm," *IEEE transactions on medical imaging,* vol. 20, no. 1, pp. 45-57, 2001.

[6] Z. Guo *et al.*, "A fast and refined cancer regions segmentation framework in whole-slide breast pathological images," *Scientific reports,* vol. 9, no. 1, pp. 1-10, 2019.

[7] S. Graham *et al.*, "Hover-net: Simultaneous segmentation and classification of nuclei in multi-tissue histology images," *Medical Image Analysis,* vol. 58, p. 101563, 2019.

[8] S. Graham *et al.*, "MILD-Net: Minimal information loss dilated network for gland instance segmentation in colon histology images," *Medical image analysis,* vol. 52, pp. 199-211, 2019.

[9] A. Mahbod *et al.*, "CryoNuSeg: A dataset for nuclei instance segmentation of cryosectioned H&E-stained histological images," *Computers in Biology and Medicine,* vol. 132, p. 104349, 2021.

[10] D. Liu *et al.*, "Unsupervised instance segmentation in microscopy images via panoptic domain adaptation and task re-weighting," in *Proceedings of the IEEE/CVF conference on computer vision and pattern recognition*, 2020, pp. 4243-4252.

[11] Q. Liang *et al.*, "Weakly supervised biomedical image segmentation by reiterative learning," *IEEE Journal of biomedical and health informatics,* vol. 23, no. 3, pp. 1205-1214, 2018.

[12] S. Hu *et al.*, "Weakly supervised deep learning for covid-19 infection detection and classification from ct images," *IEEE Access,* vol. 8, pp. 118869-118883, 2020.

[13] H. E. Atlason, A. Love, S. Sigurdsson, V. Gudnason, and L. M. Ellingsen, "Unsupervised brain lesion segmentation from MRI using a convolutional autoencoder," in *Medical Imaging 2019: Image Processing*, 2019, vol. 10949: International Society for Optics and Photonics, p. 109491H.

[14] A. Kanezaki, "Unsupervised image segmentation by backpropagation," in *2018 IEEE international conference on acoustics, speech and signal processing (ICASSP)*, 2018: IEEE, pp. 1543-1547.

[15] R. Achanta, A. Shaji, K. Smith, A. Lucchi, P. Fua, and S. Süsstrunk, "SLIC superpixels compared to state-of-the-art superpixel methods," *IEEE transactions on pattern analysis and machine intelligence,* vol. 34, no. 11, pp. 2274-2282, 2012.

[16] Y. Shen and H. Zhou, "Double dip: Re-evaluating security of logic encryption algorithms," in *Proceedings of the on Great Lakes Symposium on VLSI 2017*, 2017, pp. 179-184.

[17] J. Shi and J. Malik, "Normalized cuts and image segmentation," *IEEE Transactions on pattern analysis and machine intelligence,* vol. 22, no. 8, pp. 888-905, 2000.

[18] Y. Boykov, O. Veksler, and R. Zabih, "Fast approximate energy minimization via graph cuts," *IEEE Transactions on pattern analysis and machine intelligence,* vol. 23, no. 11, pp. 1222-1239, 2001.

[19] D. A. Clausi, "K-means Iterative Fisher (KIF) unsupervised clustering algorithm applied to image texture segmentation," *Pattern Recognition,* vol. 35, no. 9, pp. 1959-1972, 2002.

[20] Z. Ji, Y. Xia, Q. Sun, Q. Chen, D. Xia, and D. D. Feng, "Fuzzy local Gaussian mixture model for brain MR image segmentation," *IEEE Transactions on Information Technology in Biomedicine,* vol. 16, no. 3, pp. 339-347, 2012.

[21] D. Comaniciu and P. Meer, "Mean shift: A robust approach toward feature space analysis," *IEEE Transactions on pattern analysis and machine intelligence,* vol. 24, no. 5, pp. 603-619, 2002.

[22] Z. F. Knops, J. A. Maintz, M. A. Viergever, and J. P. Pluim, "Normalized mutual information based registration using k-means clustering and shading correction," *Medical image analysis,* vol. 10, no. 3, pp. 432-439, 2006.

[23] L.-H. Juang and M.-N. Wu, "MRI brain lesion image detection based on color-converted K-means clustering segmentation," *Measurement,* vol. 43, no. 7, pp. 941-949, 2010.

[24] Z. Fan, J. Lu, C. Wei, H. Huang, X. Cai, and X. Chen, "A hierarchical image matting model for blood vessel segmentation in fundus images," *IEEE Transactions on Image Processing,* vol. 28, no. 5, pp. 2367-2377, 2018.

[25] A. B. Tosun, M. Kandemir, C. Sokmensuer, and C. Gunduz-Demir, "Object-oriented texture analysis for the unsupervised segmentation of biopsy images for cancer detection," *Pattern Recognition,* vol. 42, no. 6, pp. 1104-1112, 2009.

[26] M. Caron, P. Bojanowski, A. Joulin, and M. Douze, "Deep clustering for unsupervised learning of visual features," in *Proceedings of the European Conference on Computer Vision (ECCV)*, 2018, pp. 132-149.

[27] S. Huang, Z. Kang, Z. Xu, and Q. Liu, "Robust deep k-means: An effective and simple method for data clustering," *Pattern Recognition,* vol. 117, p. 107996, 2021.

[28] L. Manduchi, K. Chin-Cheong, H. Michel, S. Wellmann, and J. Vogt, "Deep Conditional Gaussian Mixture Model for Constrained Clustering," *Advances in Neural Information Processing Systems,* vol. 34, 2021.

[29] W. Kim, A. Kanezaki, and M. Tanaka, "Unsupervised learning of image segmentation based on differentiable feature clustering," *IEEE Transactions on Image Processing,* vol. 29, pp. 8055-8068, 2020.

[30] A. Tsai, W. Wells, C. Tempany, E. Grimson, and A. Willsky, "Mutual information in coupled multi-shape model for medical image segmentation," *Medical image analysis,* vol. 8, no. 4, pp. 429-445, 2004.

[31] X. Ji, J. F. Henriques, and A. Vedaldi, "Invariant information clustering for unsupervised image classification and segmentation," in *Proceedings of the IEEE/CVF International Conference on Computer Vision*, 2019, pp. 9865-9874.

[32] X. Xia and B. Kulis, "W-net: A deep model for fully unsupervised image segmentation," *arXiv preprint arXiv:1711.08506,* 2017.

[33] M. Chen, T. Artières, and L. Denoyer, "Unsupervised object segmentation by redrawing," *arXiv preprint arXiv:1905.13539,* 2019.

[34] Y. Gandelsman, A. Shocher, and M. Irani, "" Double-DIP": Unsupervised Image Decomposition via Coupled Deep-Image-Priors," in *Proceedings of the IEEE/CVF Conference on Computer Vision and Pattern Recognition*, 2019, pp. 11026-11035.

[35] B. Zong *et al.*, "Deep autoencoding gaussian mixture model for unsupervised anomaly detection," in *International conference on learning representations*, 2018.

[36] A. Van Den Oord and B. Schrauwen, "Factoring variations in natural images with deep gaussian mixture models," in *Neural Information Processing Systems*, 2014.

[37] F. G. Zanjani, S. Zinger, B. E. Bejnordi, and J. A. van der Laak, "Histopathology stain-color normalization using deep generative models," 2018.

[38] S. Sun, Z. Cao, H. Zhu, and J. Zhao, "A survey of optimisation methods from a machine learning perspective," *IEEE transactions on cybernetics,* vol. 50, no. 8, pp. 3668-3681, 2019.

[39] K. Mardia, H. Southworth, and C. Taylor, "On bias in maximum likelihood estimators," *Journal of statistical planning and inference,* vol. 76, no. 1-2, pp. 31-39, 1999.

[40] M. Sandler, A. Howard, M. Zhu, A. Zhmoginov, and L.-C. Chen, "Mobilenetv2: Inverted residuals and linear bottlenecks," in *Proceedings of the IEEE conference on computer vision and pattern recognition*, 2018, pp. 4510-4520.

[41] W. M. Association, "World Medical Association Declaration of HelsinkiE thical Principles for Medical Research Involving Human Subjects," *Chinese Journal of Integrative Medicine,* vol. 2, no. 3, pp. 92-95, 2001.

[42] J. Wang and J. Jiang, "Unsupervised deep clustering via adaptive GMM modeling and optimisation," *Neurocomputing,* vol. 433, pp. 199-211, 2021.

[43] L. Zhou and W. Wei, "DIC: deep image clustering for unsupervised image segmentation," *IEEE Access,* vol. 8, pp. 34481-34491, 2020.

[44] O. Ronneberger, P. Fischer, and T. Brox, "U-Net: Convolutional Networks for Biomedical Image Segmentation," in *International Conference on Medical Image Computing & Computer-assisted Intervention*, 2015.

[45] E. K. Wang *et al.*, "Multi-path dilated residual network for nuclei segmentation and detection," *Cells,* vol. 8, no. 5, p. 499, 2019.

[46] X. Li, Y. Wang, Q. Tang, Z. Fan, and J. Yu, "Dual U-Net for the segmentation of overlapping glioma nuclei," *Ieee Access,* vol. 7, pp. 84040-84052, 2019.

[47] K. Tian *et al.*, "Weakly-Supervised Nucleus Segmentation Based on Point Annotations: A Coarse-to-Fine Self-Stimulated Learning Strategy," in *International Conference on Medical Image Computing and Computer-Assisted Intervention*, 2020: Springer, pp. 299-308.

[48] H. Qu *et al.*, "Weakly supervised deep nuclei segmentation using partial points annotation in histopathology images," *IEEE Transactions on Medical Imaging,* vol. 39, no. 11, pp. 3655-3666, 2020.

[49] N. Kumar, R. Verma, S. Sharma, S. Bhargava, A. Vahadane, and A. Sethi, "A dataset and a technique for generalized nuclear segmentation for computational pathology," *IEEE transactions on medical imaging,* vol. 36, no. 7, pp. 1550-1560, 2017.

[50] D. Liu, D. Zhang, Y. Song, H. Huang, and W. Cai, "Panoptic feature fusion net: a novel instance segmentation paradigm for biomedical and biological images," *IEEE Transactions on Image Processing,* vol. 30, pp. 2045-2059, 2021.

[51] D. Liu *et al.*, "Pdam: A panoptic-level feature alignment framework for unsupervised domain adaptive instance segmentation in microscopy images," *IEEE Transactions on Medical Imaging,* vol. 40, no. 1, pp. 154-165, 2020.

[52] D. Zhang *et al.*, "Exploring task structure for brain tumor segmentation from multi-modality MR images," *IEEE Transactions on Image Processing,* vol. 29, pp. 9032-9043, 2020.

[53] D. Zhang, G. Huang, Q. Zhang, J. Han, J. Han, and Y. Yu, "Cross-modality deep feature learning for brain tumor segmentation," *Pattern Recognition,* vol. 110, p. 107562, 2021.

[54] D. Zhang, J. Zhang, Q. Zhang, J. Han, S. Zhang, and J. Han, "Automatic pancreas segmentation based on lightweight DCNN modules and spatial prior propagation," *Pattern Recognition,* vol. 114, p. 107762, 2021.

[55] H.-Y. Zhou *et al.*, "Generalized Organ Segmentation by Imitating One-shot Reasoning using Anatomical Correlation," in *International Conference on Information Processing in Medical Imaging*, 2021: Springer, pp. 452-464.